\documentclass[%
 reprint,
nofootinbib,
 amsmath,amssymb,
 aps,
]{revtex4-1}

\usepackage{graphicx,color}
\usepackage{dcolumn}
\usepackage{bm}
\usepackage{lipsum}
\usepackage{amsmath}
\usepackage{float}
\usepackage{slashed}
\usepackage{graphicx}
\usepackage[compat=1.1.0]{tikz-feynman}
\usepackage{caption,ulem}
\usepackage{ragged2e}
\usepackage{subcaption}
\usepackage{xcolor}
\usepackage{siunitx}

\begin{document}

\title{Effects of fermions in one-loop propagators in the Curci–Ferrari-Delbourgo-Jarvis gauge.}

\author{Santiago Cabrera}
\affiliation{Instituto de F\'isica, Facultad de Ciencias, Universidad de la Rep\'ublica, Montevideo, Uruguay}%

\author{Marcela Pel\'aez}
\affiliation{Instituto de F\'isica, Facultad de Ingenier\'ia, Universidad de la Rep\'ublica, Montevideo, Uruguay}%

\author{Matthieu Tissier}%
\affiliation{Sorbonne Universit\'e, CNRS, Laboratoire de Physique Th\'eorique de
la Mati\`ere Condens\'ee, LPTMC, F-75005 Paris, France}%

\date{\today}

\begin{abstract}
We present the one-loop computation of the quark propagator in the Curci-Ferrari-Delbourgo-Jarvis (CFDJ) gauge, extending previous analyses to include dynamical quarks. Using the infrared-safe renormalization scheme, we study how finite gauge parameters affect the infrared behavior of QCD correlation functions. The coupling, gluon mass, and gauge parameter are found to freeze below a finite energy scale, confirming the infrared stability of the framework.  The quark dressing function $Z(p)$ shows a change in concavity between the Landau and finite-$\xi$ cases, suggesting that nonvanishing gauges may better reproduce lattice trends. 
These results establish the CFDJ gauge as a possible infrared-safe setting candidate for perturbative QCD with massive gluons. In the case of a consistency check from lattice calculations, it could provide a basis for future studies of the quark-gluon vertex and related observables. 
\end{abstract}

\maketitle

\section{Introduction}

The study of QCD correlation functions in the Landau gauge have been thoroughly investigated in the lattice in the past two decades \cite{Cucchieri:2008fc,Cucchieri:2008qm,Bogolubsky:2009dc,Boucaud:2008ky,Oliveira:2009eh,Oliveira:2012eh}. Two striking properties have been unambiguously evidenced: firstly, the gluon correlation function is screened at small momentum, and a screening mass is generated. Second, the interaction in the ghost-gluon sector remains moderate at infrared scales. In a series of studies, the Curci-Ferrari model, treated in perturbation theory (see below for details), was used successfully as an effective model to describe the correlation functions observed in lattice simulations \cite{Tissier:2010ts,Tissier:2011ey,Pelaez:2014mxa,Pelaez:2013cpa,Pelaez:2017bhh,Figueroa:2021sjm,Barrios:2021cks,Barrios:2022hzr,Barrios:2024ixj}, see \cite{Pelaez:2021tpq} for a review. The perturbative calculations indeed lead to renormalization trajectories which are finite down to the deep infrared and no Landau pole is observed. Comparisons with correlation functions obtained from the lattice show a 15$\%$ discrepancy at 1 loop and improve to few percent at two loops in Landau gauge.
The study of QCD correlation functions in covariant gauges beyond the Landau gauge is an important step toward a more complete understanding of strong-interaction dynamics. In particular, it is of great importance to understand if the two properties observed in the Landau gauge (generation of the screening mass and moderate coupling constant) are generic or if they are only valid in the Landau gauge. 

A particularly interesting class of gauge-fixings is provided by the Curci–Ferrari–Delbourgo–Jarvis (CFDJ) gauges~\cite{Curci:1976bt, Delbourgo_1982}, which are renormalizable and possess non-renormalization theorems~\cite{Wschebor:2007vh,Tissier:2008nw} that simplify the analysis of correlation functions. In particular, renormalization was studied up to five loops in $\overline{MS}$ \cite{Browne:2002wd,Gracey:2023unc,Gracey:2024oed}.
One striking feature that makes CFDJ gauges attractive is that, as for the Landau gauge, they can be regarded as an extremization problem, allowing for a direct implementation on the lattice~\cite{Serreau:2013ila}. 
This makes them promising candidates for future nonperturbative simulations in QCD with dynamical quarks.

In Ref.~\cite{Serreau:2015yna}, the gluon and ghost propagators were computed at one-loop order in the CFDJ gauge within the quenched approximation (where the fluctuations of the quarks are neglected). The aim of the present paper is to extend this program by including dynamical quarks and to perform the one-loop calculation of the full quark propagator in the CFDJ gauge. This study is motivated by several considerations. First, the excellent agreement obtained in the Landau gauge indicates that exploring other gauges can provide valuable cross-checks and insights into the issue of gauge dependence. Moreover, the CFDJ gauge shares many of the renormalization properties of the Landau gauge \cite{Wschebor:2007vh,Tissier:2008nw}, while, at the same time, being suitable for lattice implementations, which makes it an appealing framework for future comparisons \cite{Serreau:2013ila}.

A complete renormalization of the quark two-point function in the infrared-safe scheme in this gauge also constitutes a necessary step toward the computation of other important quantities, such as the quark–gluon vertex and the associated chromomagnetic moment \cite{Guzman:2025qbq,Bermudez:2017bpx}. In addition, the quark mass function in the CFDJ framework is a renormalization-group invariant: once the initial conditions for the renormalization-group flow are fixed, the emergent infrared mass scale can be directly confronted with lattice results without introducing any additional multiplicative renormalization factors. 

Another motivation stems from the fact that, in Landau gauge, the one-loop quark dressing function exhibits a convexity property that is not in agreement with lattice data \cite{Pelaez:2014mxa} \footnote{This feature is an artifact of the one loop problem: the two-loop calculation \cite{Barrios:2021cks} leads to the correct concavity.} The present calculation offers the possibility to test whether this feature persists in the CFDJ gauge, and to clarify whether it should be interpreted as an effect of the renormalization group or as a genuine gauge artifact. Finally, even in the absence of unquenched lattice simulations in the CFDJ gauge, our results provide concrete predictions for the impact of quark loops on the gluon propagator in this setting which could be compared to lattice simulation \cite{Oliveira:2026}.

This paper is organized as follows. In Sec.~\ref{sec:framework}, we review the CFDJ gauge and present the relevant Feynman rules including quarks. Sections~\ref{sec:two point functions} and ~\ref{sec:renormlization} are devoted to the one-loop calculation of the gluon, ghost and quark propagators and their renormalization. Our results are discussed in Sec.~\ref{sec:results}, with emphasis on the dependence on the gauge parameter and the implications for the quark mass function. We conclude in Sec.~\ref{sec:conclusions} with a summary and perspectives for future work.

\section{Review of the CFDJ model}
\label{sec:framework}

In this section we review the CF  model \cite{Curci:1976bt} in a four-dimensional euclidean space, including $N_f$ degenerate quarks. The gauge fixed Lagrangian is then given by:
\begin{equation}
    \mathcal{L}=\mathcal{L}_{\text{YM}}+\mathcal{L}_{\text{GF}}+\mathcal{L}_m+\mathcal{L}_{\text{matter}},
\end{equation}
where the terms correspond to the Yang-Mills, the gauge fixing, the mass and the quark  Lagrangians.

The Yang-Mills Lagrangian reads:
\begin{equation}
    \mathcal{L}_{\text{YM}}=\frac{1}{4}\left(F^{a}_{\mu\nu}\right)^2,
\end{equation}
where $F_{\mu\nu}^a=\partial_{\mu}A^{a}_{\nu}-\partial_{\nu}A^{a}_{\mu}+g_0f^{abc}A_{\mu}^{b}A_{\nu}^{c}$ is the bare field strength, $g_0$ the bare gauge coupling and $A_{\mu}^{a}$ the bare gauge field. The structure constants $f^{abc}$ are chosen completely antisymmetric.

The gauge-fixing term, expressed in terms of the Nakanishi-Lautrup field $h$ and the ghost-antighost fields $c$ and $\bar c$ is given by:
\begin{equation}
    \begin{split}
        \mathcal{L}_{\text{GF}}=&\partial_{\mu}\bar{c}^{a}\big(D_{\mu}c\big)^{a}+\frac{\xi_0}{2}h^ah^a+ih^a\partial_{\mu}A_{\mu}^{a}
        \\
    &-i\frac{\xi_0}{2}g_0f^{abc}h^a\bar{c}^{b}c^{c}-\xi_0\frac{g_0^2}{4}\big(f^{abc}\bar{c}^{b}c^{c}\big)^2.
    \end{split}
    \label{eq:lagCFDJnsym}
\end{equation}
where $\xi_0$ is the bare gauge parameter. In the limit $\xi_0\to 0$, we recover the usual Landau gauge. For $\xi_0\neq 0$, there exists a 4-ghost interaction, typical of nonlinear gauges. Note that it is possible to shift the $h$-field to obtain an action which is explicitly symmetric under the permutation of the ghost and antighost fields \cite{Tissier:2008nw}.

The effective mass term reads:
\begin{equation}
    \mathcal{L}_{m}=m_0^2\left(\frac{1}{2}\left(A_{{\mu}}^a\right)^2+\xi_0\bar{c}^{a}c^{a}\right).
    \label{eq:massLagrangian}
\end{equation}
where $m_0$ is the bare screening mass. We insist on the fact that the gluon mass appears at the level of the gauge-fixed action, contrarily to the Procca action. In this sense, gauge symmetry has already been eliminated, prior to the introduction of the mass term. The theory is renormalizable, as proven long ago~\cite{Curci:1976bt, Delbourgo_1982}. Notice that the ghosts have a mass term away from the Landau gauge ($\xi_0\neq 0$). This is imposed by symmetry considerations \cite{Curci:1976bt}.

Finally, the matter content of the theory is the usual fermionic quark Lagrangian:
\begin{equation}
    \mathcal{L}_{\text{matter}}=\sum_{i=1}^{N_f}\bar{\psi}_i\left(\slashed{\partial}+M_0^{i}\right)\psi_i,
\end{equation}
where in particular we are going to consider $M_0^i=M_0$ for all quarks. The generalization to different quark masses is straightforward.

Notice that the gauge fixing we are considering is intrinsically different from linear gauge fixing, and there is no redefinition of fields that allow us to go from one to the other. However, both models coincide in the $\xi_0\rightarrow0$ limit: they are two unequivalent extensions of the Landau gauge. 

Let us now present the Feynman rules for this model. Firstly, due to the presence of a ghost mass [see Eq.~\eqref{eq:massLagrangian}], the ghost propagator reads:
\begin{figure}[H]
    \centering
    \begin{minipage}{0.4\columnwidth}
        \raggedright
        \begin{tikzpicture}
            \begin{feynman}
                \vertex (a) {a};
                \vertex [right=2cm of a] (b) {b};
                \diagram*{
                (a) -- [charged scalar, momentum={[arrow shorten=0.2]$p$}] (b),
                };
            \end{feynman}
        \end{tikzpicture}
    \end{minipage}%
    \hspace{-0.5cm}  
    \begin{minipage}{0.4\columnwidth}
        \raggedleft
        \begin{equation}
            = \quad \frac{\delta^{ab}}{p^2 + \xi m^2 }
        \end{equation}
    \end{minipage}
\end{figure}

For determining the gluon and $h$-field propagator, we need to compute the matrix
\begin{equation}
    \mathcal M=\begin{pmatrix}
        \frac{\delta^2S}{\delta A_\mu^a\delta A_\nu^b}&        \frac{\delta^2S}{\delta A_\mu^a\delta h^b}\\
\frac{\delta^2S}{\delta h^a\delta A_\nu^b}&        \frac{\delta^2S}{h^a\delta h^b}
    \end{pmatrix}
\end{equation}
 In Fourier space, it can be expressed in terms of longitudinal and transverse projectors
\begin{align}
    P^\parallel_{\mu\nu}(p)&=\frac{p_\mu p_\nu}{p^2}\\
    P^\perp_{\mu\nu}(p)&=\delta_{\mu\nu}-\frac{p_\mu p_\nu}{p^2}
\end{align}
as:
\begin{equation}
     \mathcal M=
        \delta^{ab}
        \begin{pmatrix}
        (p^2+m_0^2)P^{\perp}_{\mu\nu}(p)+m_0^2P^{\parallel}_{\mu\nu}(p) & -p_{{\mu}} \\
        p_{\nu} & \xi_0
        \end{pmatrix},
\end{equation}
where we use the following convention for the Fourier transform:
\begin{equation}
    f(x)=\int \frac {d^dp}{(2\pi)^d}f(p)e^{i p x}
\end{equation}

We invert this matrix to obtain the bare  gluon-gluon, $h-h$ and gluon-$h$  propagators, which read 

\begin{figure}[H]
    \centering
    \hspace{-2cm}
    \begin{minipage}{0.3\columnwidth} 
        \begin{tikzpicture}
            \begin{feynman}
                \vertex (a) {a,$\mu$};
                \vertex [right=2cm of a] (b) {b,$\nu$};
                \diagram*{
                (a) -- [gluon, momentum={[arrow shorten=0.2]$p$}] (b),
                };
            \end{feynman}
        \end{tikzpicture}
    \end{minipage}%
    \hspace{0.01\columnwidth}
    \begin{minipage}{0.3\columnwidth}
        \raggedleft
        \begin{equation*}
        =\quad 
            \delta^{ab}\left( \frac{P^{\perp}_{\mu\nu}(p)}{p^2 + m_0^2 }+\frac{\xi_0 P^{\parallel}_{\mu\nu}(p)}{p^2 + \xi_0 m_0^2 }\right)
            \label{eq:propgluon}
        \end{equation*}
    \end{minipage}
\end{figure}

\begin{figure}[H]
    \centering
    \hspace{-2cm}
    \begin{minipage}{0.3\columnwidth}
        \raggedright
        \begin{tikzpicture}
            \begin{feynman}
                \vertex (a) {a};
                \vertex [right=2cm of a] (b) {b};
                \diagram*{
                (a) -- [double, momentum={[arrow shorten=0.2]$p$}] (b),
                };
            \end{feynman}
        \end{tikzpicture}
    \end{minipage}%
    \hspace{0.01\columnwidth}
    \begin{minipage}{0.3\columnwidth}
        \raggedleft
        \begin{equation*}
        =\quad
            \delta^{ab}\frac{m_0^2}{p^2 + \xi_0 m_0^2 }
        \end{equation*}
    \end{minipage}
\end{figure}

\begin{figure}[H]
    \centering
    \hspace{-2cm}
    \begin{minipage}{0.3\columnwidth}
        \raggedright
        \begin{tikzpicture}
            \begin{feynman}
                \vertex (a) {a};
                \vertex [right=2cm of a] (b) {b,$\mu$};
                \vertex [right=0.75cm of a] (c) ;
                \diagram*{
                (a) -- [double] (c),
                (c) -- [gluon] (b),
                (a) -- [scalar, opacity=0, momentum={[arrow shorten=0.2]$p$}] (b),
                };
                
            \end{feynman}
        \end{tikzpicture}
    \end{minipage}%
    \hspace{0.01\columnwidth}
    \begin{minipage}{0.3\columnwidth}
        \raggedleft
        \begin{equation*}
        = \quad 
                \delta^{ab} \frac{p_{\mu}}{p^2 + \xi_0 m_0^2 }
        \end{equation*}
    \end{minipage}
\end{figure}

As for the quark propagator, contained in the $\mathcal{L}_{\text{matter}}$ part of the Lagrangian, there is no modification to the usual Landau-one, neither from the inclusion of the mass term nor from the fact that we are working with an arbitrary $\xi-$gauge:

\begin{figure}[H]
    \centering
    \begin{minipage}{0.4\columnwidth}
        \raggedright
        \begin{tikzpicture}
            \begin{feynman}
                \vertex (a) {a};
                \vertex [right=2cm of a] (b) {b};
                \diagram*{
                (a) -- [anti fermion, momentum={[arrow shorten=0.2]$p$}] (b),
                };
            \end{feynman}
        \end{tikzpicture}
    \end{minipage}%
    \hspace{-0.5cm}  
    \begin{minipage}{0.4\columnwidth}
        \raggedleft
        \begin{equation*}
            = \quad \delta^{ab}\frac{i\slashed{p}+M_0}{p^2 + M_0^2 }
        \end{equation*}
    \end{minipage}
\end{figure}

Moving on now to the different vertices of the theory, we have the standard 3- and 4-gluon ones, as well as the gluon-ghost and gluon-quark ones, which remain unchanged from the usual Landau gauge ones. However, as explained previously, we have now to take into consideration two new ones: an $h\bar{c}c$ one and a $\bar{c}c\bar{c}c$ one for the CFDJ gauge fixing.

\begin{figure}[H]
    \centering
    \begin{minipage}{0.25\columnwidth}
        \raggedright
        \begin{tikzpicture}
            \begin{feynman}
                \vertex (a) {a};
                \vertex [below= of a] (d) ;
                \vertex [below left= of d] (c) {c} ;
                \vertex [below right= of d] (b) {b};
                \diagram*{
                (a) -- [double,momentum={[arrow shorten=0.2]$p$}] (d),
                (d) -- [charged scalar,rmomentum'={[arrow shorten=0.2]$r$}] (c),
                (b) -- [charged scalar, momentum'={[arrow shorten=0.2]$q$}] (d),
                };
                
            \end{feynman}
        \end{tikzpicture}
    \end{minipage}%
    \hspace{0.05\columnwidth}
    \begin{minipage}{0.25\columnwidth}
        \raggedleft
        \begin{equation*}
         = \quad 
            \xi_0\frac{ig_0}{2} f^{abc}
        \end{equation*}
    \end{minipage}
\end{figure}
\begin{figure}[H]
    \centering
    \begin{minipage}{0.25\columnwidth}
        \raggedright
        \begin{tikzpicture}
            \begin{feynman}
                \vertex (a) {a};
                \vertex [below=1cm of a] (g);
                \vertex [right=1cm of g] (e) ;
                \vertex [below=2cm of a] (d) {d} ;
                \vertex [right=2cm of d] (c) {c};
                \vertex [right=2cm of a] (b) {b};
                \diagram*{
                (a) -- [charged scalar] (e),
                (e) -- [charged scalar] (b),
                (e) -- [charged scalar] (c),
                (d) -- [charged scalar](e)
                };
                
            \end{feynman}
        \end{tikzpicture}
    \end{minipage}%
    \begin{minipage}{0.4\columnwidth}
      \raggedleft
        \begin{equation*}
        = \quad 
            -\xi_0\frac{g_0^2}{2} \left(f^{eab}f^{ecd}-f^{ead}f^{ecb}\right)
        \end{equation*}
    \end{minipage}
\end{figure}


\section{Two-point correlation functions}
\label{sec:two point functions}

We  parametrize the gluon and ghost two-point vertex in the standard way:
\begin{align*}
    \Gamma_{A^{a}_{\mu}A^{b}_{\nu}}^{(2)}(p)&=\delta^{ab}\left(\Gamma^{\perp}_{AA}(p)P^{\perp}_{\mu\nu}(p)+\Gamma^{\parallel}_{AA}(p)P^{\parallel}_{\mu\nu}(p)\right),
    \\
    \Gamma_{c^a\bar{c}^b }^{(2)}(p) &= \delta^{ab}\Gamma_{c\bar{c}}(p).
\end{align*}

Let us comment on the longitudinal sector of the theory. In the Landau case, the gluon two-point vertex has a contribution proportional to the mass but the gluon-gluon propagator is purely transverse (that is, proportional to $P^\perp_{\mu\nu}$, as it should in the Landau gauge. In the general case $\xi_0\neq 0$, the gluon-gluon propagator has a longitudinal part, which could be obtained from lattice simulations. In the following, we shall compute both the transverse and longitudinal parts of this propagator.

At one loop, the ghost two-point vertex has four contributions, given by the following diagrams:
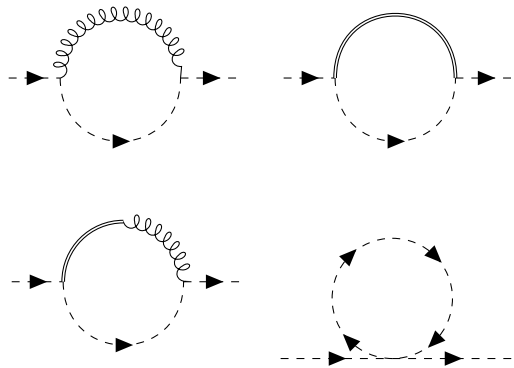
\begin{figure}[H]
     \centering
     \begin{subfigure}[b]{0.4\columnwidth}
         \centering
            \begin{tikzpicture}
            \begin{feynman}
                \vertex (a) {};
                \vertex [right=0.8cm of a] (c);
                \vertex [right=1.6cm of c] (d) ;
                \vertex [right=0.8cm of d] (b) {};
                \vertex [right=0.8cm of c] (e);
                \vertex [above=0.8cm of e] (f);
                \vertex [below=0.8cm of e] (g);
                \diagram*{
                (a) -- [charged scalar] (c) -- [gluon, half left, looseness=1.8] (d) --[charged scalar](b),
                (c) --[charged scalar, half right, looseness=1.8](d)
                };
                
            \end{feynman}
\end{tikzpicture}
         \caption*{}
         \label{fig:selfGhost_a}
     \end{subfigure}
     \hspace{-0.02cm}
      \begin{subfigure}[b]{0.4\columnwidth}
         \centering
            \begin{tikzpicture}
            \begin{feynman}
                \vertex (a) {};
                \vertex [right=0.8cm of a] (c);
                \vertex [right=1.6cm of c] (d) ;
                \vertex [right=0.8cm of d] (b) {};
                \vertex [right=0.8cm of c] (e);
                \vertex [above=0.8cm of e] (f);
                \vertex [below=0.8cm of e] (g);
                \diagram*{
                (a) -- [charged scalar] (c) -- [double, half left, looseness=1.8] (d) --[charged scalar](b),
                (c) --[charged scalar, half right, looseness=1.8](d)
                };
                
            \end{feynman}
\end{tikzpicture}
         \caption*{}
         \label{fig:selfGhost_b}
     \end{subfigure}
     \begin{subfigure}[b]{0.4\columnwidth}
         \centering
        \begin{tikzpicture}
            \begin{feynman}
                \vertex (a) {};
                \vertex [right=0.8cm of a] (c);
                \vertex [right=1.6cm of c] (d) ;
                \vertex [right=0.8cm of d] (b) {};
                \vertex [right=0.8cm of c] (e);
                \vertex [above=0.8cm of e] (f);
                \vertex [below=0.8cm of e] (g);
                \diagram*{
                (a) -- [charged scalar] (c) -- [double, quarter left] (f) --[gluon, quarter left](d) --[charged scalar](b),
                (c) --[charged scalar, half right, looseness=1.8](d)
                };           
            \end{feynman}
\end{tikzpicture}
         \caption*{}
         \label{fig:selfGhost_c}
     \end{subfigure}
     \hspace{-0.1cm}
     \begin{subfigure}[b]{0.4\columnwidth}
         \centering
         \begin{tikzpicture}
            \begin{feynman}
                \vertex (a) {};
                \vertex [right=0.8cm of a] (c);
                \vertex [right=1.6cm of c] (d) ;
                \vertex [right=0.8cm of d] (b) {};
                \vertex [right=0.8cm of c] (e);
                \vertex [above=0.8cm of e] (f);
                \vertex [above=0.8cm of f] (g);
                \vertex [right=0.8cm of f] (h);
                \vertex [left=0.8cm of f] (i);
                \diagram*{
                (a) -- [charged scalar] (e) -- [charged scalar, quarter left] (i) --[charged scalar, quarter left] (g) -- [charged scalar, quarter left] (h) --
                [charged scalar, quarter left](e) -- [charged scalar] (b)
                };              
            \end{feynman}
\end{tikzpicture}
         \caption*{}
         \label{fig:selfGhost_d}
     \end{subfigure}
        \caption{Diagrams contributing to the ghost self energy at one loop.}
        \label{fig:selfGhost}
\end{figure}

Notice that only the first one remains in the Landau limit $\xi_0\to 0$. It is important to note that the fourth diagram, corresponding to a ghost tadpole, is not identically zero in dimensional regularization because we are considering a massive propagator for the ghost in an arbitrary CFDJ gauge. Should we not have included the $m_0^2\xi_0\bar{c}^{a}c^{a}$ term in the action, this diagram would not contribute to the two-point function. 

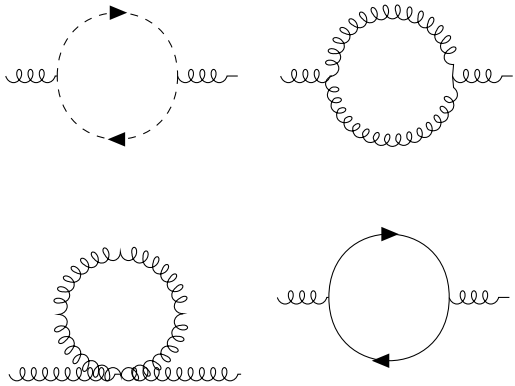
\begin{figure}[H]
     \centering
     \begin{subfigure}[b]{0.4\columnwidth}
         \centering
            \begin{tikzpicture}
            \begin{feynman}
                \vertex (a) {};
                \vertex [right=0.8cm of a] (c);
                \vertex [right=1.6cm of c] (d) ;
                \vertex [right=0.8cm of d] (b) {};
                \vertex [right=0.8cm of c] (e);
                \vertex [above=0.8cm of e] (f);
                \vertex [below=0.8cm of e] (g);
                \diagram*{
                (a) -- [gluon] (c) -- [charged scalar, half left, looseness=1.8] (d) --[gluon](b),
                (d) --[charged scalar, half left, looseness=1.8](c)
                };
                
            \end{feynman}
\end{tikzpicture}
         \caption*{}
         \label{fig:selfGluon_a}
     \end{subfigure}
     \hspace{-0.02cm}
      \begin{subfigure}[b]{0.4\columnwidth}
         \centering
            \begin{tikzpicture}
            \begin{feynman}
                \vertex (a) {};
                \vertex [right=0.8cm of a] (c);
                \vertex [right=1.6cm of c] (d) ;
                \vertex [right=0.8cm of d] (b) {};
                \vertex [right=0.8cm of c] (e);
                \vertex [above=0.8cm of e] (f);
                \vertex [below=0.8cm of e] (g);
                \diagram*{
                (a) -- [gluon] (c) -- [gluon, half left, looseness=1.8] (d) --[gluon](b),
                (c) --[gluon, half right, looseness=1.8](d)
                };
                
            \end{feynman}
\end{tikzpicture}
         \caption*{}
         \label{fig:selfGluon_b}
     \end{subfigure}
     \hspace{-0.1cm}
     \begin{subfigure}[b]{0.4\columnwidth}
         \centering
         \begin{tikzpicture}
            \begin{feynman}
                \vertex (a) {};
                \vertex [right=0.8cm of a] (c);
                \vertex [right=1.6cm of c] (d) ;
                \vertex [right=0.8cm of d] (b) {};
                \vertex [right=0.8cm of c] (e);
                \vertex [above=0.8cm of e] (f);
                \vertex [above=0.8cm of f] (g);
                \vertex [right=0.8cm of f] (h);
                \vertex [left=0.8cm of f] (i);
                \diagram*{
                (a) -- [gluon] (e) -- [gluon, quarter left] (i) --[gluon, quarter left] (g) -- [gluon, quarter left] (h) --
                [gluon, quarter left](e) -- [gluon] (b)
                };
                
            \end{feynman}
\end{tikzpicture}
         \caption*{}
         \label{fig:selfGluon_c}
     \end{subfigure}
     \begin{subfigure}[b]{0.4\columnwidth}
         \centering
            \begin{tikzpicture}
            \begin{feynman}
                \vertex (a) {};
                \vertex [right=0.8cm of a] (c);
                \vertex [right=1.6cm of c] (d) ;
                \vertex [right=0.8cm of d] (b) {};
                \vertex [right=0.8cm of c] (e);
                \vertex [above=0.8cm of e] (f);
                \vertex [below=0.8cm of e] (g);
                \diagram*{
                (a) -- [gluon] (c) -- [fermion, half left, looseness=1.8] (d) --[gluon](b),
                (c) --[anti fermion, half right, looseness=1.8](d)
                };
                
            \end{feynman}
\end{tikzpicture}
         \caption*{}
         \label{fig:selfGluon_d}
\end{subfigure}
        \caption{Diagrams contributing to the gluon self energy at one loop.}
        \label{fig:selfGluon}
\end{figure}

 The diagrams contributing to the gluon self energy are the same as those appearing (see Fig.~\ref{fig:selfGluon}) in the Landau gauge but must be computed with a massive ghost propagator and a longitudinal gluon propagator.

\begin{figure}[H]
    \centering
        \begin{tikzpicture}
            \begin{feynman}
                \vertex (a) {};
                \vertex [right=1cm of a] (c);
                \vertex [right=1.6cm of c] (d) ;
                \vertex [right=1cm of d] (b) {};
                \vertex [right=0.8cm of c] (e);
                \vertex [above=0.8cm of e] (f);
                \vertex [below=0.8cm of e] (g);
                \diagram*{
                (a) -- [gluon] (c) -- [charged scalar, half left, looseness=1.7] (d) --[double](b),
                (d) --[charged scalar, half left, looseness=1.7](c)
                };
            \end{feynman}
        \end{tikzpicture}
            \begin{tikzpicture}
                \begin{feynman}
                    \vertex (a) {};
                    \vertex [right=1cm of a] (c);
                    \vertex [right=1.6cm of c] (d) ;
                    \vertex [right=1cm of d] (b) {};
                    \vertex [right=0.8cm of c] (e);
                    \vertex [above=0.8cm of e] (f);
                    \vertex [below=0.8cm of e] (g);
                    \diagram*{
                    (a) -- [double] (c) -- [charged scalar, half left, looseness=1.7] (d) --[double](b),
                    (d) --[charged scalar, half left, looseness=1.7](c)
                    };
            \end{feynman}
        \end{tikzpicture}       
    \caption{Contributions at one loop to the $h^{a}A^{b}_{\mu}$ vertex and $h^{a}$-field self energy, respectively.}
    \label{fig:autoenergiaH}
\end{figure}
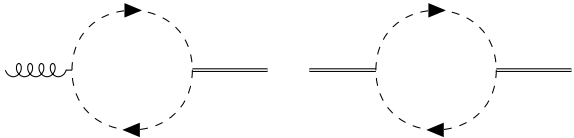

 We finally consider the self energies involving the $h$ field that we parametrize as:
\begin{align*}
    \Gamma_{h^{a}A^{b}_{\mu}}(p) &= - \Gamma_{A^{a}_{\mu}h^{b}}(p) =\delta^{ab}p_{\mu}\Gamma_{hA}(p),
    \\
    \Gamma_{h^{a}h^{b}}(p) &=\delta^{ab}\Gamma_{hh}(p).
\end{align*}
 At one loop, there is one diagram contributing to each of these, see Fig.~\ref{fig:autoenergiaH},

Moving now to the matter content of the model, we parametrize the two-point quark function by two scalar functions $Z(p)$ and $M(p)$, defined by:
\begin{equation}
    \Gamma_{\psi\bar{\psi}}^{(2)}(p)=Z^{-1}(p)(i\slashed{p}+M(p)).
    \label{eq:paramFunQQ}
\end{equation}

At one loop, this quark-quark vertex involves only a single diagram, given by:
\begin{figure}[H]
    \centering
    \centering
        \begin{tikzpicture}
            \begin{feynman}
                \vertex (a) {};
                \vertex [right=1cm of a] (c);
                \vertex [right=1.6cm of c] (d) ;
                \vertex [right=1cm of d] (b) {};
                \vertex [right=0.8cm of c] (e);
                \vertex [above=0.8cm of e] (f);
                \vertex [below=0.8cm of e] (g);
                \diagram*{
                (a) -- [fermion] (c) -- [gluon, half left, looseness=1.7] (d) --[fermion](b),
                (d) --[anti fermion, half left, looseness=1.7](c)
                };
        \end{feynman}
    \end{tikzpicture}
    \caption{Diagram contributing to the quark self energy at one loop.}
    \label{fig:selfQuark}
\end{figure}
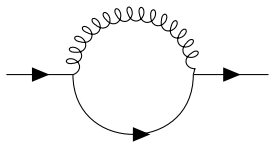

All these diagrams, after substitution of the vertices and propagators, can be reduced to linear combinations of Passarino-Veltman master integrals, $A(m)$ and $B(p,m_1,m_2)$:
\begin{align}
    A(m) &= \int\frac{d^dq}{(2\pi)^d} \frac{1}{q^2+m^2},
    \\
    B(p,m_1,m_2) &=\int\frac{d^dq}{(2\pi)^d} \frac{1}{(q+p)^2+m_1^2}\frac{1}{q^2+m_2^2},
\end{align}
which are UV divergent and need to be regularized. To this end, we use dimensional regularization. The divergences appear through poles in $\epsilon=(4-d)/2$. We shall explain the renormalization scheme implemented in the following section. 

\section{Renormalization}
\label{sec:renormlization}
\subsection{Renormalization scheme}

Let us now study the one-loop renormalization of this theory in $d=4$, part of which was already done in a previous work~\cite{Serreau:2015yna}, without the presence of quarks nevertheless. As already mentioned, all the diagrams presented until now are UV divergent. This problem can be addressed by the usual means of renormalization group techniques. We are going to define the renormalized fields in the standard way, introducing renormalization factors, as follows:

\begin{equation}
    \begin{aligned}
        A^{a,\mu}     &= \sqrt{Z_A}\, A_R^{a,\mu},       &\quad
        c^a           &= \sqrt{Z_c}\, c_R^a, \\
        \bar{c}^a     &= \sqrt{Z_c}\, \bar{c}_R^a,       &\quad
        h^a           &= \sqrt{Z_h}\, h_R^a, \\
        \psi          &= \sqrt{Z_\psi}\, \psi_R.
    \end{aligned}
\end{equation}

and similarly for the couplings:
\begin{equation}
    \begin{matrix}
        m_0^2 = Z_{m^2}m_R^2, && g_0 = Z_g g_R, \\ \xi_0 = Z_AZ_{\xi}^{-1}\xi_R, && M_0=Z_M M_R.
    \end{matrix}
\end{equation}

In order to fix the finite part of the renormalization factors, we must choose a renormalization scheme. In the present work, we are going to consider the infrared safe scheme (IS) which has been extensively used in the framework of the CF model~\cite{Tissier:2010ts,Tissier:2011ey,Pelaez:2014mxa,Pelaez:2013cpa,Barrios:2021cks,Barrios:2020ubx,Barrios:2022hzr}, giving rise to well-behaved propagators for the gluon deep into the infrared, in agreement with lattice calculations.  It is given by:
    \begin{align}
        \label{eq:renScheme1}
         \Gamma_{AA}^{R\perp}(p=\mu) &= \mu^2+m_R^2,
          \hspace{1mm}
          \Gamma_{c\bar{c}}^R(p=\mu) = \mu^2+\xi_R m_R^2, 
         \nonumber\\
         \Gamma_{hh}^R(p=\mu) &= \xi_R, 
         \hspace{1mm}
         \Gamma_{hA}^R(p=\mu) = 1,
         \nonumber\\
         Z(p=\mu) &= 1, \hspace{1mm}
         M(p=\mu) = M_R,
    \end{align}

and
\begin{subequations}         
    \begin{align}
         Z_{\xi}^2 &= Z_cZ_AZ_{m^2}, 
         \\
         Z_g &= Z_{\xi}^2Z_c^{-1}Z_A^{-1/2}.
    \end{align}
    \label{eq:renScheme2}
\end{subequations}
The last two conditions are due to non-renormalization theorems for the mass and the interaction coupling in the CF model, proven in~\cite{Wschebor:2007vh}.
Strictly speaking, these relations concern the divergent part of the renormalization factors. In the IS scheme, we extend these relations, imposing them to the finite parts as a means to fix them. Observe that the theory can be renormalized by considering 2-point vertex function, even to fix the renormalization factor of the interaction. This major  simplification generalizes what can be done in the Landau gauge when using the Taylor scheme  \cite{Taylor:1971ff} and is a consequence of the nonrenormalization theorems \eqref{eq:renScheme2}.

From now on, except otherwise stated, all fields, coupling constants and masses are the renormalized ones and we shall not put the $R$ to alleviate the notations. At one loop, the renormalization conditions lead to the following expressions for the renormalization factors, defining $Z_{\alpha}=1+\delta Z_{\alpha}$. 
    
    \begin{align}
        \delta Z_{\xi} &= 2\Gamma_{hA}^{1\text{lo}}+\frac{\Gamma_{hh}^{1\text{lo}}}{\xi},
         \hspace{1mm}
        \delta Z_{c} = -\frac{\xi m^2}{\mu^2}\delta Z_{\xi}-\frac{\Gamma^{1\text{lo}}_{\bar{c}c}}{\mu^2},
        \nonumber\\
        \delta Z_{A} &= \frac{m^2}{\mu^2}(\delta Z_{c}-\delta Z_{\xi})-\frac{\Gamma_{AA}^{\perp1\text{lo}}}{\mu^2},
      \hspace{1mm}
        \delta Z_{h} = -\delta Z_{A}+2\Gamma_{hA}^{1\text{lo}},
        \nonumber\\
        \delta Z_{m^2} &= 2 \delta Z_{\xi}-\delta Z_{A}-\delta Z_{c},  \hspace{1mm}
        \delta Z_g = 2\delta Z_{\xi}-\delta Z_{c}-\frac{1}{2}\delta Z_{A},
        \nonumber\\
        \delta Z_{\psi} &= -Z^{-1,1\text{lo}}, \hspace{1mm}
        \delta Z_{M} = Z^{-1,1\text{lo}}\left(1-\frac{M^{1\text{lo}}}{M}\right).
    \end{align}
Here, all the functions are evaluated at the renormalization scale (i.e. $\Gamma=\Gamma(p=\mu)$) and $M^{1\text{lo}}$ corresponds to the parametrization \eqref{eq:paramFunQQ} of the one-loop result for the quark-quark vertex.

\subsection{Renormalization Group}

The renormalization scheme described above leads to finite expressions for the two point functions. However, as usual, these involve "large logarithms", when evaluated at scales $p$ much bigger than the renormalization scale $\mu$. In order to cure this shortcoming of the method, we employ the renormalization group equations, which enables us to relate the vertices renormalized at two, arbitrarily separated, scales.

The idea is to evaluate the perturbative expression at momenta which are always close to the renormalization scale $\mu$. To do so, we introduce a sliding scale $\mu$ and study how the theory varies when we change this otherwise arbitrary scale $\mu$. This can be done by deriving the beta functions and $z$ normalization parameters which encode the variation of the parameters of the theory (masses, coupling constant, gauge parameter) and the field normalization change when the renormalization scale is modified.
In particular, once we renormalize at a certain scale $\mu_0$ the various vertices, we are interested in obtaining the relationship between their value at the arbitrarily chosen renormalization scale $\mu_0$ and at a sliding scale $\mu$. In order to do so, we use the general renormalization group equation for the vertex functions. For the cases we are interested, the RG equation gives us for the vertices we are interested in:
\begin{align}
\Gamma_{\phi_1\phi_2}^R(p,\mu)&=\sqrt{z_{\phi_1}(\mu;\mu_0)z_{\phi_2}(\mu;\mu_0)}\Gamma_{\phi_1\phi_2}^R(p,\mu_0),
\end{align}
where
\begin{align}
    \log z_\phi(\mu) &= \int_{\mu_0}^{\mu}\frac{d\mu'}{\mu'}\gamma_\phi(\mu')
    \label{eq:zchicos}
\end{align}
$\phi$ represents any of the field of the theory,
and the vertices renormalized at $\mu_0$ are given by simple expressions, the ones appearing in the renormalization scheme \eqref{eq:renScheme1}. In \eqref{eq:zchicos}, we introduced the anomalous dimensions of the fields, which are defined, together with the $\beta$-functions as:
\begin{align}
    \beta_X &= \mu\left.\frac{dX_R}{d\mu}\right\vert_{g_0,m_0,\xi_0,M_0}
\end{align}
where $X=\{g_R,m_R,\xi_R,M_R\}$, and,
\begin{align}
    \gamma_{\phi} &= \mu\left.\frac{d \log Z_{\phi}}{d\mu}\right\vert_{g_0,m_0,\xi_0,M_0} 
\end{align}
where once again $\phi$ represents any of the field of the theory.
For some of the anomalous dimensions and $\beta-$functions, we can deduce identities starting from the non-renormalization theorems \eqref{eq:renScheme2}. Firstly, the $\beta-$function for the gauge parameter $\xi$ is obtained as:
\begin{equation}
    \beta_{\xi}=\mu\frac{\partial\xi}{\partial\mu}=\mu\frac{\partial}{\partial\mu}\left(\frac{Z_\xi}{Z_A}\xi_B\right)=\xi\left(\gamma_\xi-\gamma_A\right)=\xi\left(-2\gamma_h\right),
    \label{eq:betaxi}
\end{equation}
where we defined $\gamma_\xi$ in the same fashion as the (actual) anomalous dimensions.

Likewise, for the mass and gauge coupling we can deduce:
\begin{equation}
    \beta_{g}=g\left(\gamma_c+\frac{\gamma_A}{2}-2\gamma_\xi\right)
    \label{eq:betag}
\end{equation}
and
\begin{equation}
    \beta_{m^2}=m^2\left(\gamma_c+\gamma_A-2\gamma_\xi\right).
    \label{eq:betaM}
\end{equation}
We can solve \eqref{eq:betaxi}-\eqref{eq:betaM} for $\gamma_c$, $\gamma_A$ and $\gamma_h$, obtaining:

    \begin{align}
        \gamma_c &= 2\frac{\beta_\xi}{\xi}-2\frac{\beta_g}{g}+3\frac{\beta_{m^2}}{m^2},\nonumber
        \\
        \gamma_A &= 2\left(-\frac{\beta_g}{g}+\frac{\beta_{m^2}}{m^2}\right),\nonumber
        \\
        \gamma_h &= -\frac{1}{2}\frac{\beta_\xi}{\xi}.
    \end{align}
Our explicit expressions are given in \cite{mathematicafile}, we also check that  UV behaviour matches with \cite{Browne:2002wd}.

Plugging this results back into the definition of the $z$ functions, we are left with simple power laws for the $z(\mu;\mu_0)$ functions:

    \begin{align}
        z_c(\mu;\mu_0) &= \frac{\xi^2(\mu)}{\xi^2(\mu_0)}\frac{g^2(\mu_0)}{g^2(\mu)}\frac{m^6(\mu)}{m^6(\mu_0)},\nonumber
        \\
        z_A(\mu;\mu_0) &= \frac{g^2(\mu_0)}{g^2(\mu)}\frac{m^4(\mu)}{m^4(\mu_0)},\nonumber
        \\
        z_h(\mu;\mu_0) &= \sqrt{\frac{\xi(\mu_0)}{\xi(\mu)}}.
    \end{align}

Unfortunately, no such simplification occurs for the quarks, and so a numerical integration of $\gamma_{\psi}(\mu)$ is required to obtain $z_{\psi}$:
\begin{equation}
    z_{\psi}(\mu) = \exp \int_{\mu_0}^{\mu}\frac{d\mu'}{\mu'}\gamma_{\psi}(\mu').
\end{equation}

\section{Results}
\label{sec:results}
\subsection{Choosing the parameters}
\label{ssec:parameters}
The $\beta$ functions obtained in the previous section can now be integrated. In this section, we discuss how we should initialize the flow. Monte-Carlo simulations are performed on a lattice, with an interaction strength which fixes the energy scale, through the dimensional transmutation phenomenon. To make contact with our analytic calculation, we have to associate a value of the coupling constant at some momentum scale $\mu_0$, expressed in GeV. At this same scale, a lattice simulation of the CFDJ would use a certain value of the $\xi$ parameter. We should initialize the gauge parameter to this same value at scale $\mu_0$. Finally, the quark mass at scale $\mu_0$ is also a parameter of the lattice simulation and we should initialize $M$ at this value. In what concerns the initialization of the coupling constant, we can use the fact that the $\beta$-function is universal in the UV (that is, independent of the gauge parameter $\xi$) to argue that if we initialize the RG flow at a sufficiently large scale $\mu_0$, this initialization can be taken independent of the parameter $\xi$. In our analysis, we therefore initialize our RG flows sufficiently deep in the UV and discard all dependence of the coupling constant $g(\mu_0)$ with the gauge-fixing parameter $\xi(\mu_0)$.

The situation is a bit more complex for the parameter $m$. As explained in the introduction, the CF model is seen as a phenomenological model to describe the low-energy limit of QCD. Following this philosophy, the initial value of $m$ must be fixed by choosing the value which best fits some result obtained by some other method, particularly, lattice calculations. Once this value is obtained, the model is predictive. In practice, the model is still predictive because we aim at reproducing several functions, with only one fitting parameter. However, in the case of the CFDJ model considered in this article, since no lattice simulation have been performed so far, we have no way of fixing the initial value of the mass parameter $m$. In what follows, for illustrative purposes we use the value of the parameter which gives the best fits in the Landau gauge. However, it is our expectation that, when compared to lattice calculations, the best fit parameter shall depend on the gauge parameter $\xi$. Therefore, we will explore a second method in the last subsection of the results: fixing $m(\mu_0;\xi(\mu_0))$ so as to obtain the same IR saturation values for M while varying $\xi(\mu_0)$.

\subsection{Gauge fixing effect on the couplings and the gauge sector}
\label{sect:constant_mass}
\begin{figure}[htpb]
    \centering
    \includegraphics[width=\columnwidth]{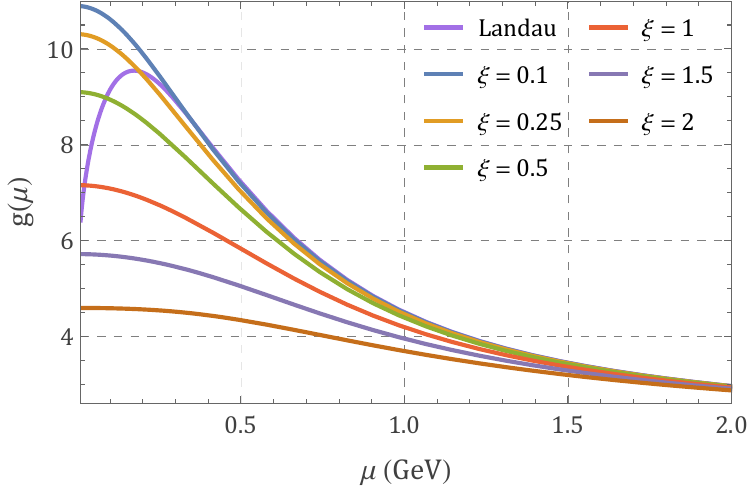}
    \includegraphics[width=\columnwidth]{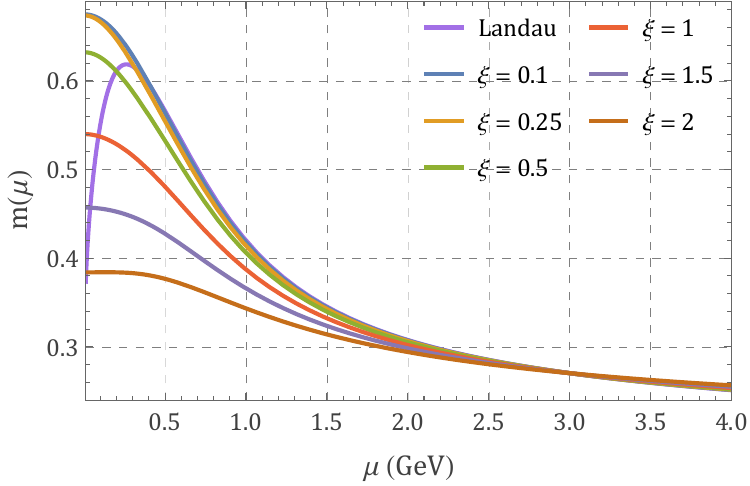}
    \includegraphics[width=\columnwidth]{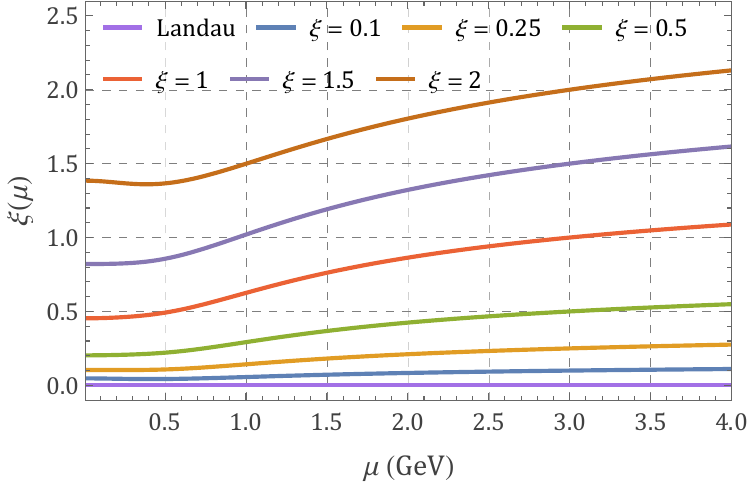}
    \caption{Running of the gauge coupling $g(\mu)$, the gluon mass $m(\mu)$ and the gauge parameter $\xi(\mu)$ in the IS scheme, for $N_f=2$ degenerate quarks.}
    \label{fig:runningsAt1GeV}
\end{figure}

In this section we present the evolution of coupling constant, the gluon mass and the gauge parameter, as well as the gluon and ghost propagators for different $\xi(\mu=\mu_0)$, in a similar way to \cite{Serreau:2015yna} but taking into consideration dynamical quarks.
In order to do so, we need to fix the values of the parameters at an arbitrary scale $\mu_0$ which must chosen as well. 
Aiming at a clear picture of the IR properties, we chose to renormalize at a scale of $\mu_0=\SI{3}{\giga\electronvolt}$. To obtain values for $g$, $m$ and $M$, we initialized the flow at $\SI{1}{\giga\electronvolt}$ for the values which provide the best fit for the Landau gauge \cite{Pelaez:2014mxa} and followed the evolution of Landau theory up to the desired scale of $\SI{3}{\giga\electronvolt}$. At this point, having no lattice data to say anything about non-zero $\xi$ values, we chose to fix the parameters, for all considered $\xi$ values, as those corresponding to the Landau case at this scale: $g=2.5$, $m=\SI{0.27}{\giga\electronvolt}$ and $M=\SI{76}{\mega\electronvolt}$.

In Fig.~\ref{fig:runningsAt1GeV} we present the results for $g(\mu)$, $m(\mu)$ and $\xi(\mu)$ which closely resemble the results in the quenched approximation \cite{Serreau:2015yna}\footnote{In order to make an exact comparison, we plotted as well our results with renormalization scale $\mu_0=\SI{1}{\giga\electronvolt}$, the one used at that work, and obtained very similar curves once dynamical quarks are taken into consideration.} These plots show that the flows freeze below a certain scale. This comes as no surprise, since all modes are massive. Consequently, when the running scale $\mu^2$ gets smaller than the (running masses squared) $m^2(\mu)$ and $\xi(\mu)m^2(\mu)$, the $\beta$ functions tend to zero. Therefore, neither $g$ nor $m$ vanish in the IR, instead going to constant values. We emphasize however, that it is not obvious \textit{a priori} that this regime is reached. The gauge fixing parameter behaves regularly at this limit reaching also a saturation value. This is to be expected, since the addition of new massive degrees of freedom, the quarks, will not change the deep IR behaviour, where all but the massless particles decouple. The only different case is that of Landau gauge, where the ghost remains massless thus preventing the RG flow from freezing. Naturally, considering massive dynamical quarks does not change this fact, and we obtain a similar picture to the pure gauge case too \cite{Serreau:2015yna}. As for the differences, we see that in the unquenched case the saturation values of the gluon mass and the coupling constant are smaller than in the quenched case, while the gauge fixing parameter freezes at higher values for all $\xi(\mu_0)\neq0$ cases.

\subsection{Ghost and gluon propagator}

\begin{figure}[H]
    \centering
    \includegraphics[width=\columnwidth]{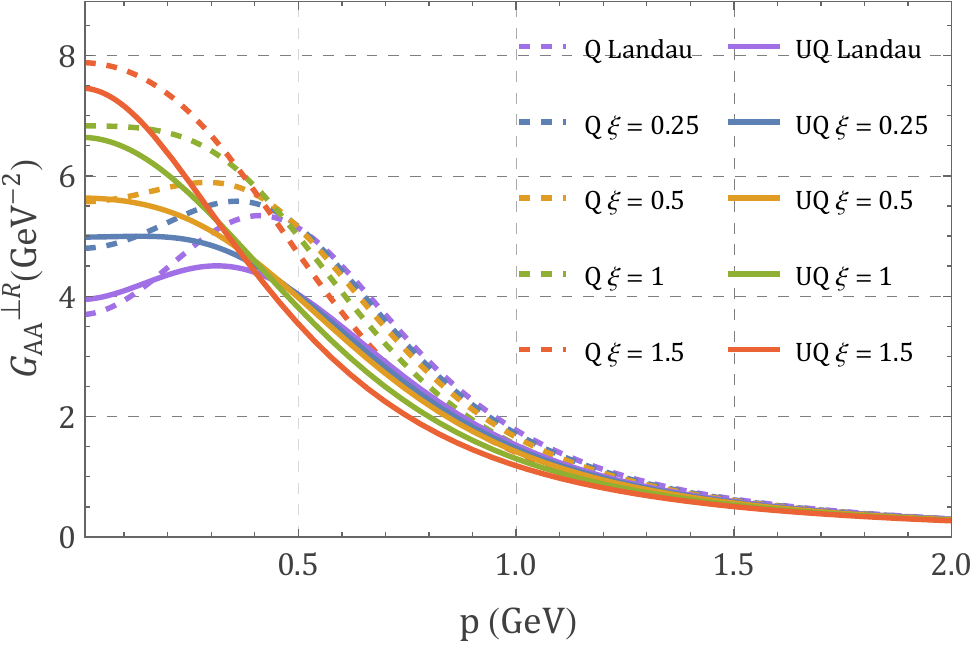}
    \includegraphics[width=\columnwidth]{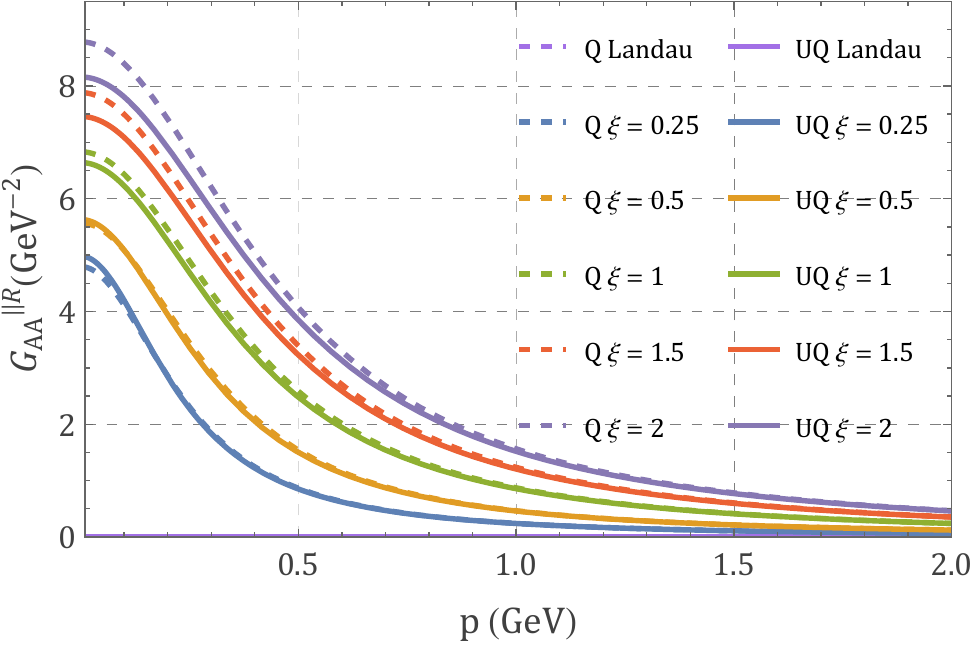}
    \includegraphics[width=\columnwidth]{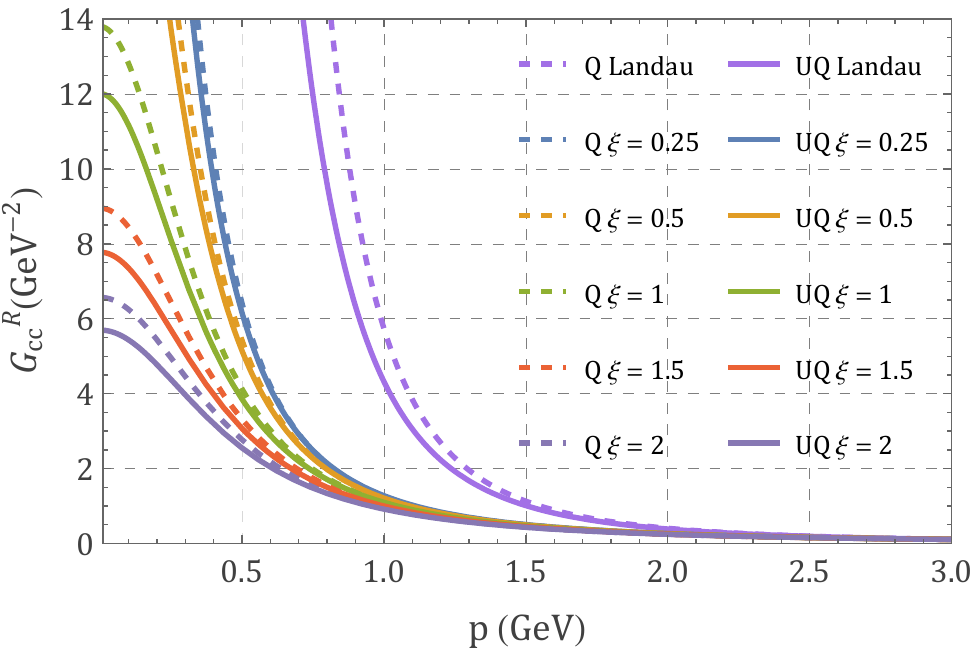}
 
    \caption{Transverse and longitudinal components of the gluon propagator and ghost propagator as functions of momentum, in both the quenched (dashed lines) and unquenched ($N_f=2$, solid lines) cases.}
    \label{fig:propGluonGhost}
\end{figure}

In this section we analyze the impact of quark fluctuations on the gauge sector, comparing the gluon and ghost propagators in the presence of $N_f=2$ degenerate quarks with the results for CFDJ gauge in the quenched approximation. We present our results for the gluon transversal and longitudinal components and for the ghost propagator in Fig~\ref{fig:propGluonGhost}. In order to obtain a clearer picture of the IR properties, we chose from this point onward to renormalize at a higher scale of $\mu_0=\SI{3}{\giga\electronvolt}$. To obtain values for $g$, $m$ and $M$, we initialized the flow at $\SI{1}{\giga\electronvolt}$ for the values which provide the best fit for the Landau gauge \cite{Pelaez:2014mxa} -- the ones we used in the previous section -- and followed the flow of Landau theory up to the new scale of $\SI{3}{\giga\electronvolt}$. At this point, having no lattice data to say anything about non-zero $\xi$ values, we chose to fix the parameters, for all considered $\xi$ values, as those corresponding to the Landau case at this scale: $g=2.5$, $m=\SI{0.27}{\giga\electronvolt}$ and $M=\SI{76}{\mega\electronvolt}$. 

The inclusion of quarks does not change the qualitative picture greatly. Notice that if one increases $\xi(\mu_0)$, the theory goes from non-monotonous to monotonous propagators. Likewise, $\xi(\mu_0)$ increases, for the particular way of fixing the parameters at $\mu_0$ we are exploring, we observe that the transverse propagator in presence of quarks goes from saturating above to below its corresponding quenched version.

As for the longitudinal component of the gluon propagator, we notice only slight quantitative changes when including quarks, while the global behaviour remains unchanged. In a similar fashion, for the ghost sector, the propagator changes very slightly after the inclusion of quarks, with the exception of the Landau case, and in particular, saturates at very similar values. This happens even though the coupling constant $g$, the mass $m$ and the gauge parameter $\xi$ have nontrivial differences in their RG flows when comparing the quenched and unquenched cases. All these effects seem to compensate for the ghost propagator.

\subsection{Quark propagator}

\subsubsection{Dressing function}

We now address the matter sector of the model. We present our results for the quark propagator for the $N_f=2$ degenerate case, once again showing results corresponding to a renormalization point at $\mu_0=\SI{3}{\giga\electronvolt}$. In this section we shall discuss the dressing function $Z(p)$ of the quark, firstly considering strict perturbation theory, and then including RG flow corrections. 

The quark dressing function has proven to be a very interesting quantity. Indeed, in \cite{Pelaez:2014mxa}, it was observed that in the Landau gauge at one loop, the dressing function fails to correctly reproduce the numerical simulation data, exhibiting a concavity opposite to the expected one. This is due to the fact that the one-loop contribution is abnormally small (in fact, it vanishes when the gluon mass is taken to zero). The two-loop contribution is therefore the dominant one.\footnote{This feature persists in analyses including ladder resummations \cite{Pelaez:2017bhh,Pelaez:2020ups}}. As expected, this issue is resolved once two-loop corrections are included \cite{Barrios:2021cks}. In this section, we show that this is a pecularity of the Landau gauge which is not present for sufficiently large gauge parameter.

\begin{figure}[H]
    \centering
        \includegraphics[width=\columnwidth]{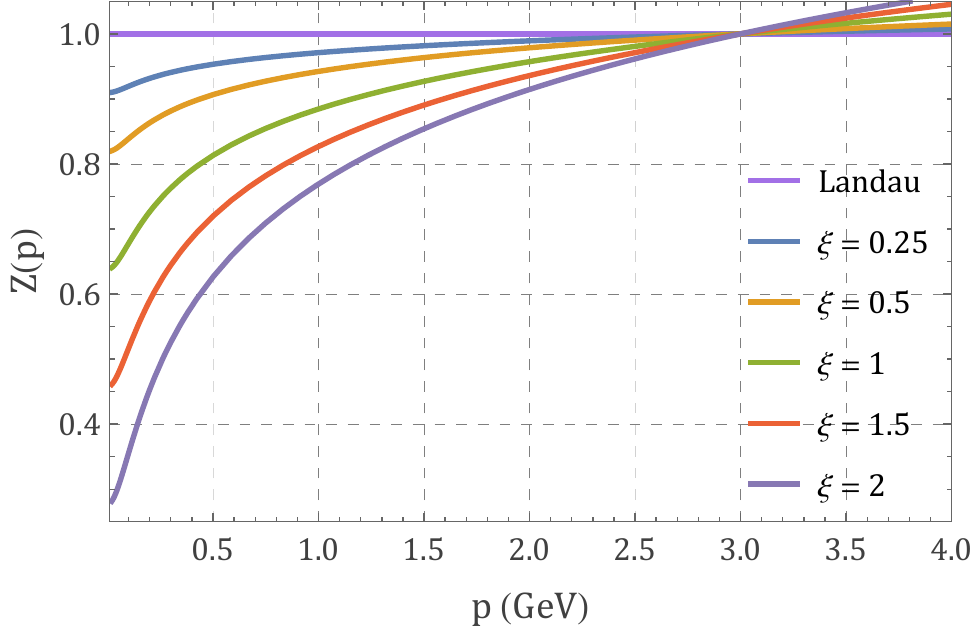}
 \includegraphics[width=\columnwidth]{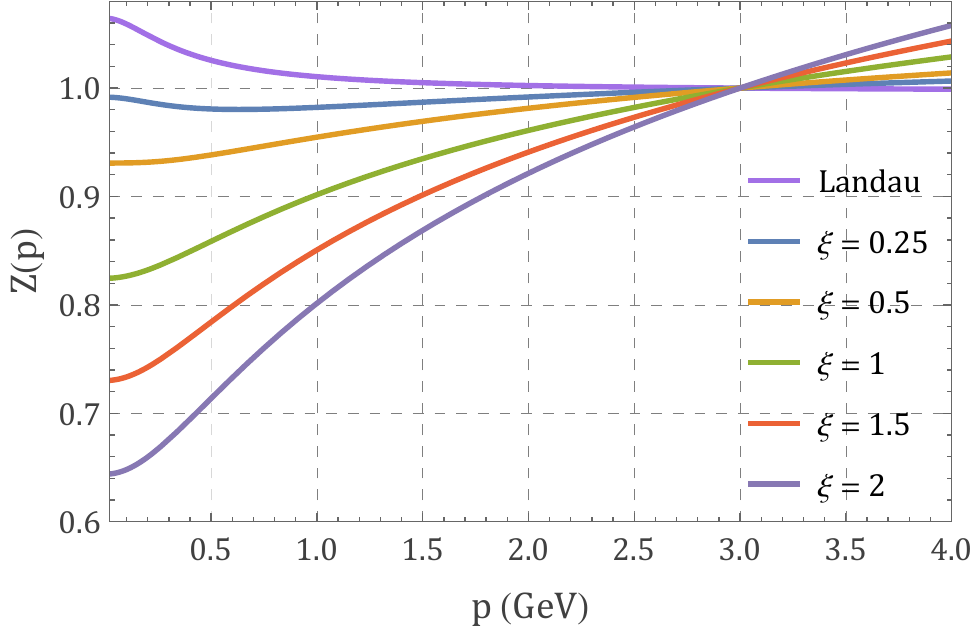}

   \caption{Quark dressing function without considering RG flow corrections, for massless gluons (top) and massive gluons (bottom), for $N_f=2$ degenerate quarks. }
    \label{fig:dressingWORG}
\end{figure}

We now discuss the properties of the dressing function $Z(p)$ from strict perturbation theory, without including renormalization group effects, in order to analyze the origin of the different behaviors.  Let us first consider the case of massless gluons ($m=0$). In the Landau gauge ($\xi=0$) the quark dressing function is constant at this order of perturbation theory and becomes an increasing function for nonvanishing parameter $\xi$, see upper curve of Fig.~\ref{fig:dressingWORG}. When the gluon mass is set to a finite value (using again $m=\SI{0.27} {\giga\electronvolt}$ at $\mu_0=\SI{3}{\giga\electronvolt}$), the dressing function is now decreasing at  $\xi=0$, and becomes an increasing function for $\xi>0.5$. For intermediate values, it is non-monotonous. This is to be compared with the result obtained in lattice simulation. In the Landau gauge, the dressing function is seen to be an increasing function \cite{Oliveira:2018lln}. It would be very interesting to compare the 1-loop predictions for the dressing function at finite $\xi$ with lattice results to see of the behavior observed in the Landau gauge is generic or exceptional.

\begin{figure}
    \centering
    \includegraphics[width=\columnwidth]{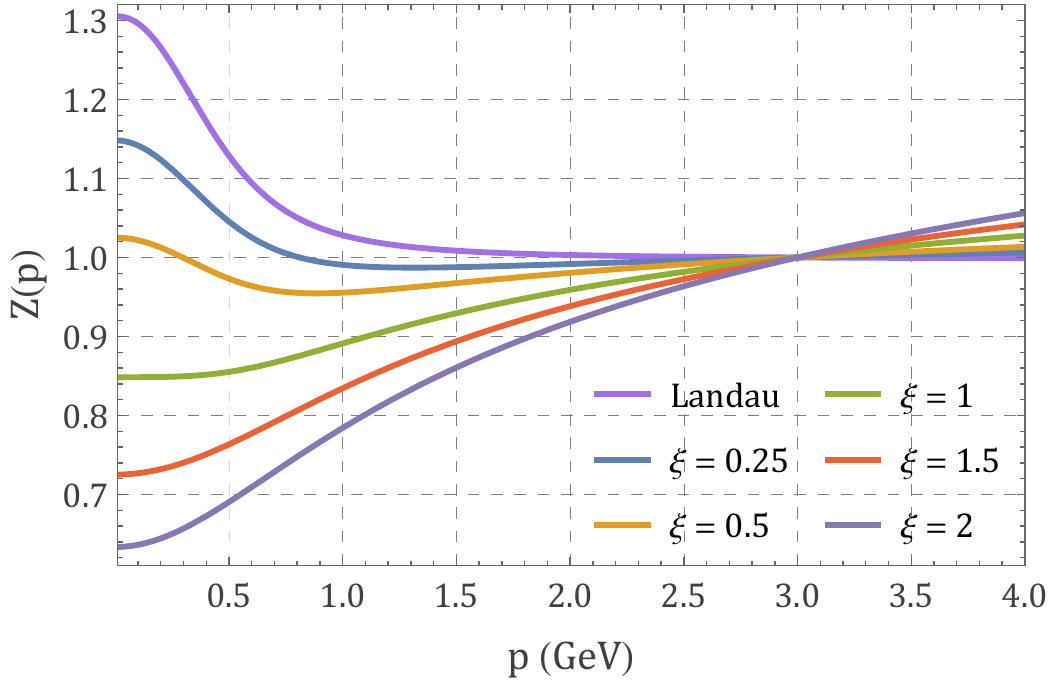}
    \caption{Quark dressing function with RG flow corrections for $N_f=2$ degenerate quarks. }
    \label{fig:dressingWRG}
\end{figure}

In Fig.~\ref{fig:dressingWRG} we present the dressing function for the quarks including RG flow corrections for the massive case, let us recall that the massless case leads to Landau poles. The qualitative picture is similar to the strict perturbation theory result, although the saturation values at the IR are further away, up and down, from 1. We again observe that for small $\xi(\mu_0)$ values, the $Z$ function grows towards the IR, while it decreases for larger gauge parameter values. Once again, it would be of great interest to compare our results with lattice calculations, to better understand the difference in concavity near the IR.  If numerical simulations were to show that, for finite gauge parameters, the concavity of the dressing function follows the same trend as observed in the Landau gauge, then we could again infer that the non-vanishing gauge case provides a better reproduction of the values obtained from the numerical simulations.

\subsubsection{Mass function}

\begin{figure}[htpb]
    \centering
    \includegraphics[width=\columnwidth]{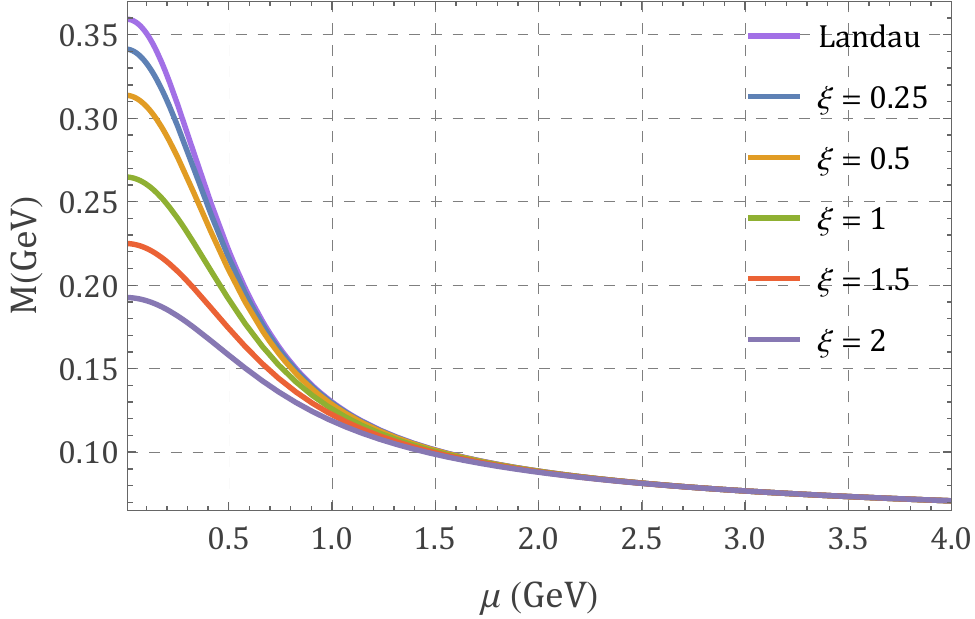}
    \caption{Running of the quark mass, initialized at $\mu_0=\SI{3}{\giga\electronvolt}$ to $\SI{76}{\mega\electronvolt}$ for various initial gauge fixing parameters $\xi(\mu_0)$ for $N_f=2$ degenerate quarks.
    \label{fig:runningQuarkMass}}
\end{figure}

In this section, for completeness and illustrative purposes, we present the behaviour of the (degenerate) quark mass, fixing its value at $\mu_0$ to $\SI{76}{\mega\electronvolt}$ as already mentioned. We present our results in Fig.~\ref{fig:runningQuarkMass}. The most interesting feature we observe is a monotonous decrease of the constituent mass of the quarks as the $\xi(\mu_0)$ value is increased. The two extreme values we considered, corresponding to Landau gauge and $\xi=4$, differ in a factor of order two ($\SI{0.36}{\giga\electronvolt}$ and $\SI{0.19}{\giga\electronvolt}$ respectively). This clearly shows the back-reaction of the gauge parameter on the quark mass.

The strong dependence of the constituent quark mass with the gauge parameter is troubling because it is at odds with our intuition that this quantity has a physical meaning and should therefore be gauge independent. 
Let us recall that these results were obtained under the hypothesis that the gluon mass at the initialization point is independent of $\xi_0$ which is questionable. In the next section, we propose a scheme that avoids this dilemma.

\subsection{Parameters fixing to match the quark mass functions}

In this last section of our results, we discuss an alternative way of fixing the parameters of the theory at the renormalization scale $\mu_0=\SI{3}{\giga\electronvolt}$.

In principle, we should fix the fitting parameter $m(\mu_0)$ so as to reproduce some property. {Since there is up to now no lattice result available in this gauge an option would be to use a physical information to fix it. We can think for instance at the quark pole mass, which is a gauge-invariant quantity. This quantity is unfortunately difficult to extract from our calculation and we will use as a proxy the constituent quark mass, that is the quark mass at vanishing momentum [$M(p=0)$]. The strategy we propose here is the following. First consider a quark in the Landau gauge, with some UV mass $M(\mu_0)$ (as before we use $M(\SI{3}{\giga\electronvolt})=\SI{76}{\mega\electronvolt}$) and compute the associated constituent quark mass $M(0)$. Assuming that this quantity is no too affected by the choice of the gauge fixing, we now choose the {\it gluon} mass at $\mu_0$ such that, for finite values of the gauge parameter, the constituent quark mass coincides with that found in the Landau gauge.

For the adjustment of $m(\mu_0)$ we implemented a straightforward bijection method until an accuracy comparable to that of the couplings RG flow numerical integration we implemented. The gluon mass decreases almost linearly from $\SI{0.27}{\giga\electronvolt}$ at $\xi =0$ to $\SI{0.2}{\giga\electronvolt}$ at $\xi =2$. \begin{figure}[htpb]
    \centering
    \includegraphics[width=\columnwidth]{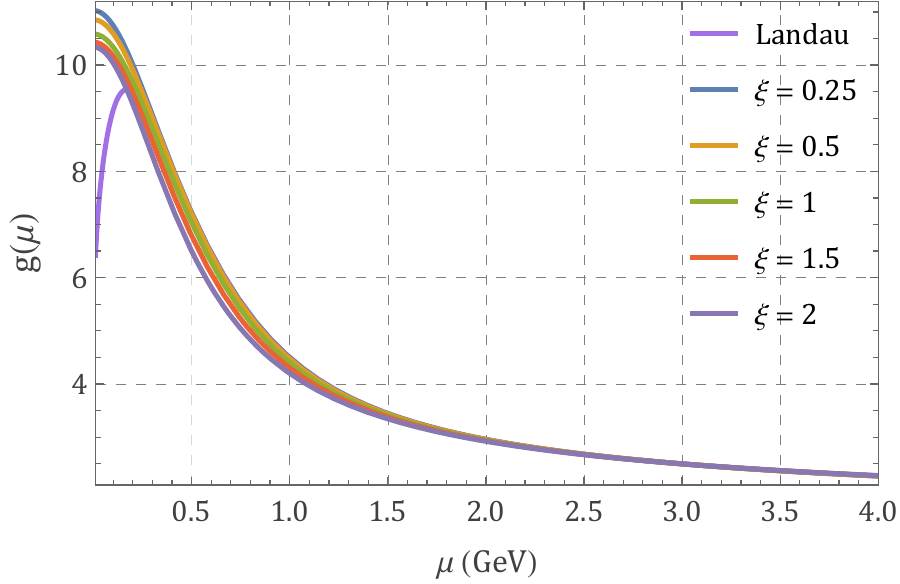}
    \includegraphics[width=\columnwidth]{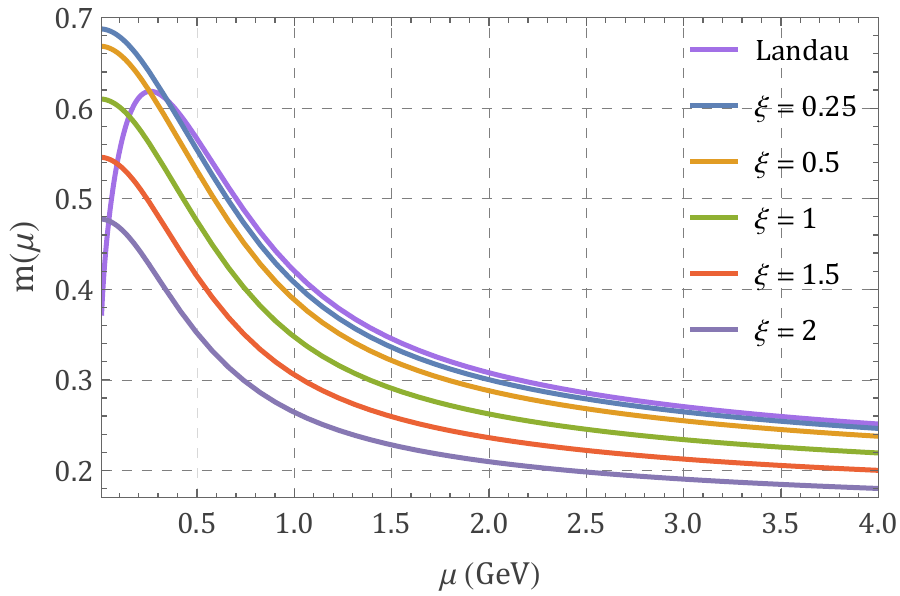}
    \includegraphics[width=\columnwidth]{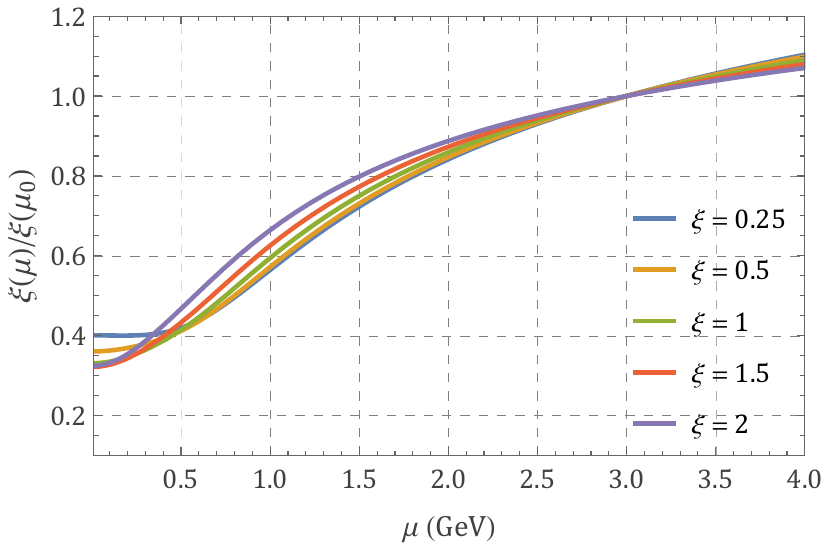}
    \caption{Running of the gauge coupling $g(\mu)$, the gluon mass $m(\mu)$ and the gauge fixing parameter $\xi(\mu)$ with the IS scheme,  for $N_f=2$ degenerate quarks, fixing $m(\mu_0)$ to have coincident $M(0)$ values, see text.}
    \label{fig:runningsFit}
\end{figure}
We present in Fig.~\ref{fig:runningsFit} our results for the coupling, the gluon mass and the gauge parameter divided by its value at $\mu_0$. We notice a few striking properties. The general trend  is that the dependence on the gauge parameter $\xi_0$ is much smaller than in the scheme discussed in Sect.~\ref{sect:constant_mass}}.  Firstly, up to the freezing scale $\mu^2\sim \xi m^2$, the couplings for all $\xi(\mu_0)$ values run in a very similar way with overlapping curves. Secondly, the gluon mass -- up to a constant -- behaves also qualitatively similar as $\xi(\mu_0)$ is changed. Finally, the gauge parameter (divided by its initial value at $\mu_0$) runs without a strong dependence of the choice of $\xi(\mu_0)$ -- all curves behave qualitatively similar, for the considered cases.

\begin{figure}
    \centering
    \includegraphics[width=\columnwidth]{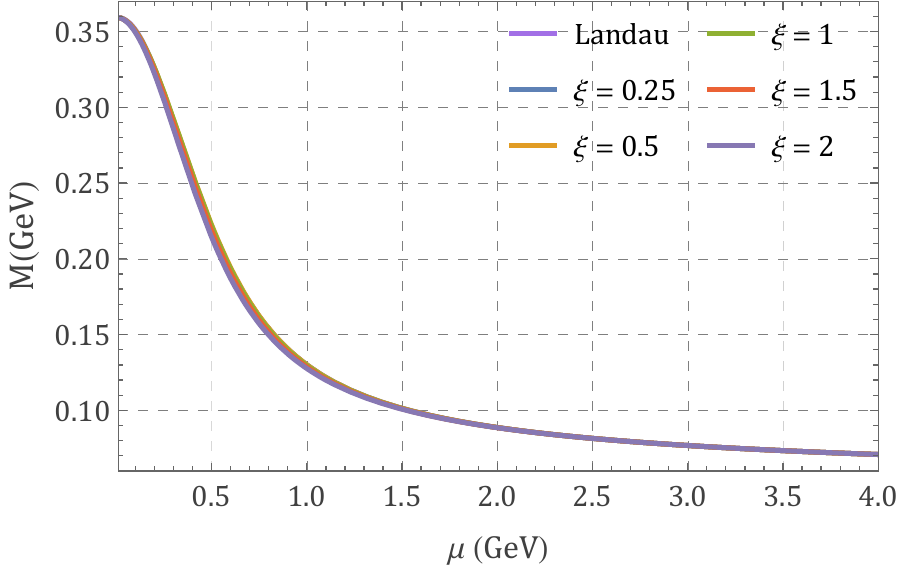}
\includegraphics[width=\columnwidth]{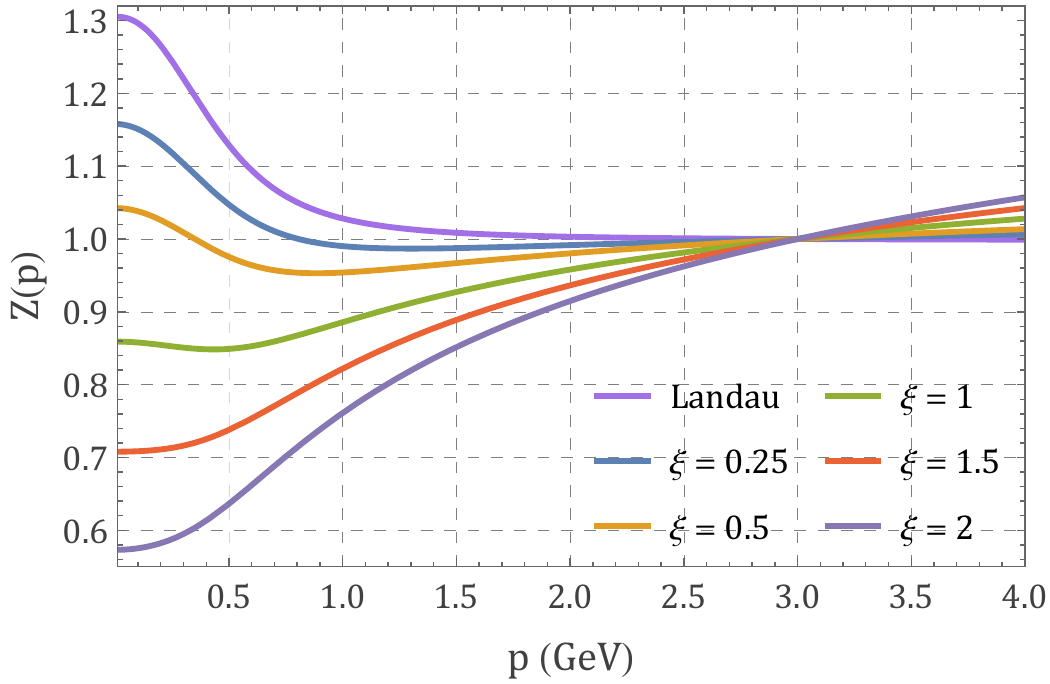}

    \caption{Quark mass  running and quark dressing function for a UV quark mass fixed at $\mu_0=\SI{3}{\giga\electronvolt}$ to $\SI{76}{\mega\electronvolt}$, for various initial gauge fixing parameters $\xi(\mu_0)$ for $N_f=2$ degenerate quarks. We chose $m(\mu_0)$ such that the constituent quark mass $M(0)$ is $\xi$-independent.
\label{fig:runningQuarkMassDressingFit}}
\end{figure}
We show in Fig.~\ref{fig:runningQuarkMassDressingFit} the quark mass functions for different gauge parameters. By construction, all the curves coincide at the UV initialization scale $\mu_0$ and at vanishing momentum. The overlap of the curves for intermediates momenta is remarkable. As for the dressing function, little differences where observed with respect to our previous analysis. We present the results in Fig.~\ref{fig:runningQuarkMassDressingFit}.

\begin{figure}[htpb]
    \centering
    \includegraphics[width=\columnwidth]{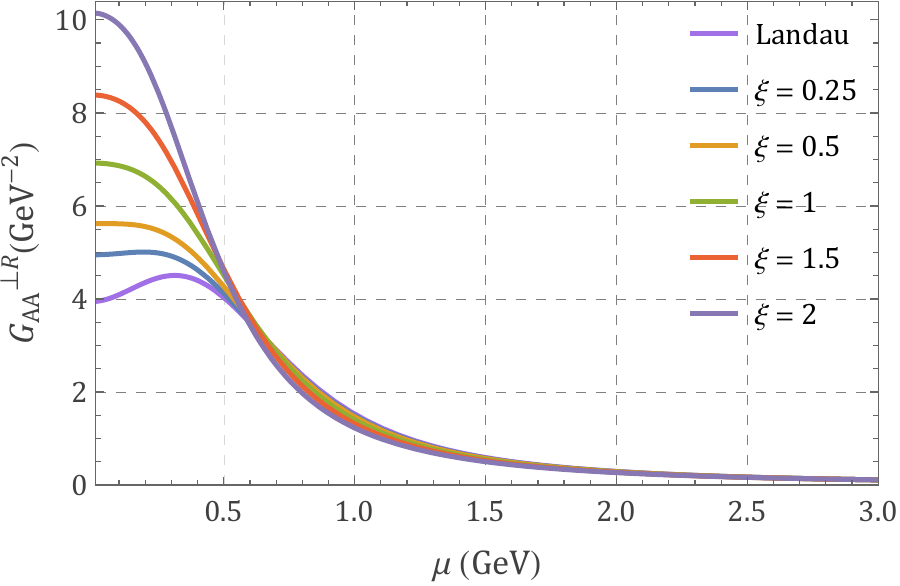}
    \includegraphics[width=\columnwidth]{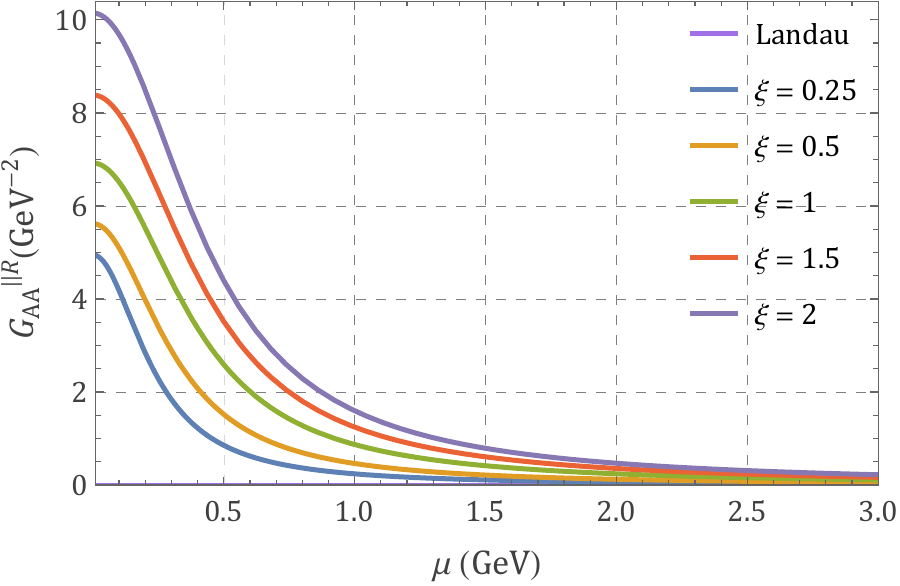}

    \includegraphics[width=\columnwidth]{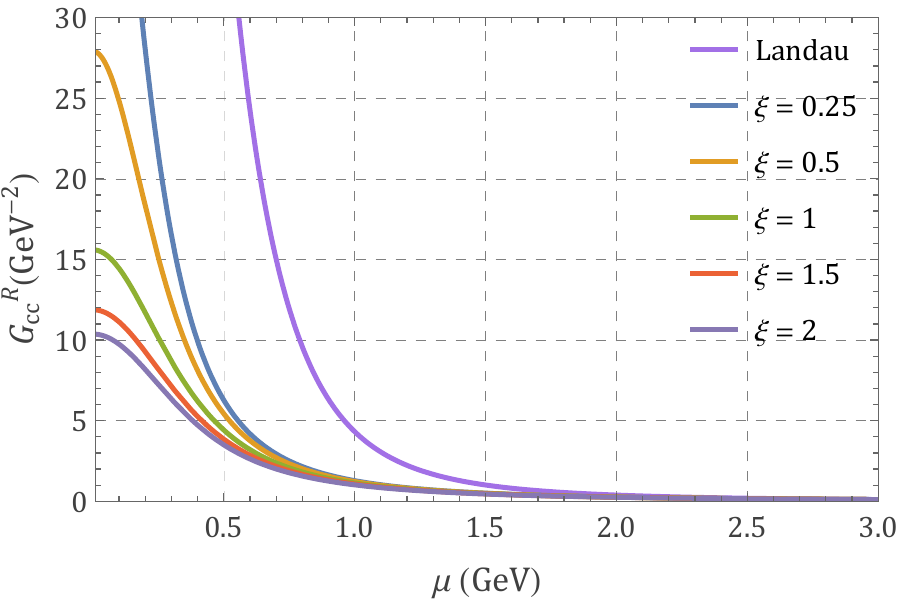}
    \caption{Transverse and longitudinal components of the gluon propagator and ghost propagator, as functions of momentum with the IS scheme for $N_f=2$ degenerate quarks. We chose $m(\mu_0)$ such that the constituent quark mass $M(0)$ is $\xi$-independent.}
    \label{fig:propsFitMass}
\end{figure}

In Fig.~\ref{fig:propsFitMass}, we show the behavior of the ghost and gluon propagators for different gauge parameters. These curves resemble those obtained in the other scheme, see Fig.~\ref{fig:propGluonGhost}. This indicates that the physical quantities are rather insensitive to the choice of scheme.

\begin{figure}
    \centering
    \includegraphics[width=\columnwidth]{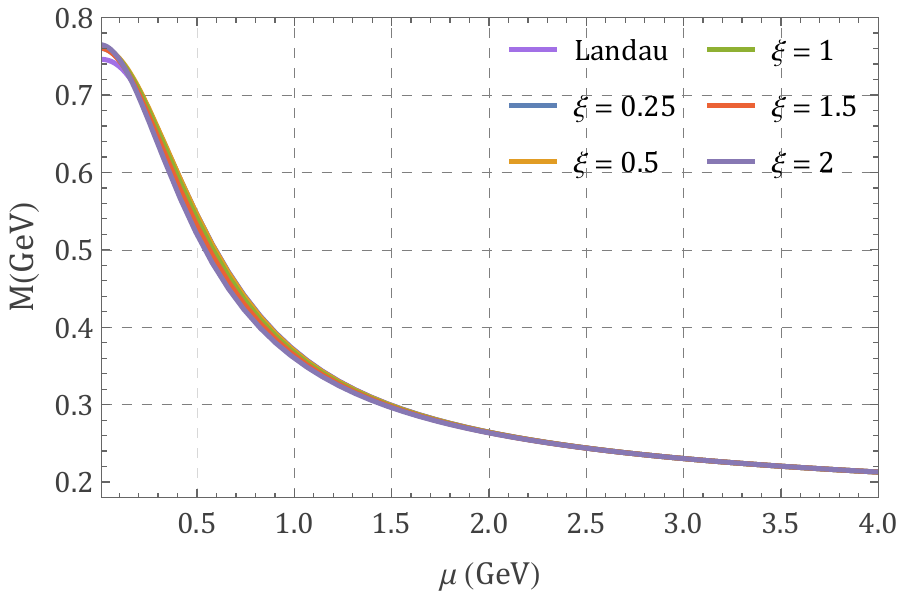}
    \caption{Running quark mass, fixed at $\mu_0=\SI{3}{\giga\electronvolt}$ to $\SI{0.23}{\giga\electronvolt}$ for various initial gauge fixing parameters $\xi(\mu_0)$ for $N_f=2$ degenerate quarks. We used the same values of  $m(\mu_0)$ as for the light quark case.
    \label{fig:runningQuarkMassFitFactor3}}
\end{figure}

To test the consistency of the method we now study the robustness of our choice of the gluon mass. To do so, we have considered another quark whose UV mass is three times as big as the previous one. We initialized all other parameters to the values obtained to generate Fig.~\ref{fig:runningsFit}. We present the result for this new quark mass function in Fig.~\ref{fig:runningQuarkMassFitFactor3}. Surprisingly, the saturation values for $M(\mu)$ differ in this case only by 2.4\% which indicates that our criterion for fixing the gluon mass is rather insensitive to our choice of the UV quark mass, which is a good signal.

\section{Conclusions}
\label{sec:conclusions}

In this work, we have computed the one-loop quark propagator in the Curci-Ferrari-Delbourgo-Jarvis (CFDJ) gauge, including the effects of dynamical quarks. Our study extends previous analyses performed in the quenched approximation and allows us to investigate how finite gauge parameters modify the infrared behavior of the correlation functions within this massive extension of QCD.  
We have verified that, similarly to the pure Yang-Mills case, the inclusion of quarks does not qualitatively change the infrared dynamics: the running coupling, gluon mass, and gauge parameter all freeze at finite values in the deep infrared. The presence of quarks slightly reduces the saturation value of the gluon mass and coupling, while increasing that of the gauge parameter.  
For the gluon and ghost propagators, the unquenched results display modest quantitative differences with respect to the quenched case, the most notable one being a moderate enhancement of the gluon propagator in the infrared region.

The analysis of the quark dressing function $Z(p)$ suggests that one-loop calculations for  finite-$\xi$ gauges are more satisfactory than in the Landau case. Indeed, the one-loop contribution to the dressing function is expected to be larger than the two-loop effect for large enough $\xi_0$, at odds with the Landau case. Therefore, we expect that our findings would compare well with lattice results.
In what concerns the quark mass function $M(p)$, we find that it decreases monotonically with increasing gauge parameter under the assumption that the initial value of the gluon mass $m(\mu_0)$ is independent of the gauge parameter. We proposed another scheme where $m(\mu_0)$ is chosen such that the constituent quark mass is gauge-independent. The ghost and gluon propagators, as well as the quark dressing function, behave similarly in these two schemes.

\section{Acknowledgments} 
We thank Orlando Oliveira, Urko Reinosa, Julien Serreau, Paulo Silva, and Nicolás Wschebor for very useful discussions related to this work. We acknowledge the financial support from the Program for the Development of Basic Sciences (PEDECIBA), the ECOS program, and ANII through the FCE\_2025\_186497. S. Cabrera acknowledges the financial support from the Comisión Académica de Posgrado (CAP).

\newpage
\bibliographystyle{unsrt}
\bibliography{references}

\end{document}